\documentclass[showpacs,preprintnumbers,amsmath,amssymb,axodraw,nofootinbib]{revtex4}

\usepackage{amsmath}
\usepackage{graphicx}
\usepackage{epsf}
\usepackage{psfrag}
\usepackage{epsfig}
\usepackage{graphics}
\usepackage[dvips]{color}
\newcommand{\be}{\begin{equation}}
\newcommand{\ee}{\end{equation}}
\newcommand{\bq}{\begin{eqnarray}}
\newcommand{\eq}{\end{eqnarray}}

\begin{document}

	\title{Residual gauge-invariance in a massive Lorentz-violating extension of QED}

	\date{\today}
	\author{B. Alves Marques$^{(a,b)}$} \email[] {bruno.alves@ifmg.edu.br}
	\author{A. P. Ba\^eta Scarpelli$^{(a)}$} \email[] {scarpelli@cefetmg.br}
	\author{J. C. C. Felipe$^{(c)}$} \email[] {jean.cfelipe@ufvjm.edu.br}
	\author{L. C. T. Brito$^{(d)}$} \email[]{lcbrito@ufla.br}
	
	\affiliation{(a)Centro Federal de Educa\c{c}\~ao Tecnol\'ogica - MG \\
		Avenida Amazonas, 7675 - 30510-000 - Nova Gameleira - Belo Horizonte
		-MG - Brazil}
	
	\affiliation{(b) Instituto Federal de Minas Gerais - Campus Sabará \\
		Rodovia MGC 262, Sobradinho, Minas Gerais, Brazil}
	
	\affiliation{(c) Instituto de Engenharia, Ci\^encia e Tecnologia, Universidade Federal dos Vales do Jequitinhonha e Mucuri \\
		Avenida Um, no. 4050, 39447-814, Cidade Universit\'aria, Jana\'uba, MG, Brazil}
	
	\affiliation{(d) Universidade Federal de Lavras - Departamento de F\'{\i}sica \\
		Caixa Postal 3037, 37.200-000, Lavras, Minas Gerais, Brazil}

	\begin{abstract}
		\noindent
		We reassess an alternative CPT-odd electrodynamics obtained from a Palatini-like procedure. Starting from a more general situation, we analyze the physical consistency of the model for different values of the parameter introduced in the mass tensor. We show that there is a residual gauge-invariance in the model if the local transformation is taken to vary only in the direction of the Lorentz-breaking vector. This residual gauge-invariance can be extended to all models whose only source of gauge symmetry breaking is such a mass term.
		
	\end{abstract}
	

	\maketitle
	
	\section{Introduction}
	It has been more than two decades that the investigation of Lorentz-violating models got the attention of the community of quantum field theory physicists. Indeed, it is believed that the usual symmetries of special relativity will be broken in the limit of very high energy physics as an effect of quantum gravity issues \cite{qgrav1,qgrav2}. This possibility was first discussed in the papers of Kosteleck\'y and Samuel \cite{qgrav1}, but it gained great dimension when Carroll, Field and Jackiw proposed a modification of the classical electrodynamics by the inclusion of a Chern-Simons-like term in the photon sector \cite{CFJ}. This term violates both CPT and Lorentz symmetries and gives rise to a wide range of interesting physical effects, which were intensively studied. Many papers treated features of the model like its physical consistency \cite{CFJ1}, the possibility of the CFJ term being radiatively generated \cite{CFJ2} and many other aspects \cite{CFJ3,CFJ4}.
	
	Further, a general description of Lorentz violation in quantum field theories was provided by the Standard Model Extension (SME) \cite{SME1}-\cite{SME3}. The SME, which includes the Carroll-Field-Jackiw (CFJ) term, establishes a set of constant tensors as the parameters of Lorentz violation, whose small magnitudes are strongly constrained by experiments \cite{data}. These constant tensors would arise from spontaneous Lorentz symmetry-breaking at very high energies \cite{SME1}. It is important to note that the SME preserves $SU(3) \times SU(2) \times U(1)$ gauge symmetry and the renormalizability.
	
	One aspect that has been gaining increasing attention is the inclusion of massive photons in models that violate Lorentz symmetry. In general, it is believed that the fact that photons are massless is a direct consequence of gauge invariance of quantum electrodynamics (QED). Nevertheless, many studies of the physics beyond the Maxwell electromagnetism have been performed, mainly in the context of Proca model \cite{Proca1}-\cite{Proca5}. Although studies using astrophysical sources and laboratory experiments put very strong constraints in the photon mass \cite{mass-limit-1}-\cite{mass-limit-7}, it may be possible that it has a very tiny but nonvanishing rest mass $m_\gamma$. It is interesting to note that the usual CFJ model already accommodates an effective mass for the photon which is proportional to the absolute value of the Lorentz-breaking vector \cite{Bonetti-1}, \cite{Bonetti-2}. Still in the context of electrodynamics with the Carroll-Field-Jackiw term, an explicit mass term for the photon was used to repair unitarity problems for a time-like background vector \cite{Manoel-Helayel}.
	
	On the other hand, models with Lorentz-violating (LVI) mass terms present remarkable peculiarities, like the ones pointed out in \cite{LVmass1} and \cite{LVmass2}, in which the gauge field has two massive degrees of freedom but the static force between charged particles is Coulomb-like. Many aspects like the quantum induction of Lorentz-breaking mass terms for the photon \cite{LVmass3}, the construction of a Stueckelberg Lagrangian in a generalized $R_\xi$ gauge \cite{LVmass4}, spontaneous gauge symmetry breaking in a Lorentz-violating gauge-Higgs model \cite{LVmass5} were investigated. In \cite{LVmass6}, some aspects of this kind of gauge-symmetry breaking were focused in a study of dual models.
	
	An alternative way of generating Lorentz-violating mass terms was investigated in \cite{Palatini}, in which a Palatini-like formulation of the CFJ model was performed. Interestingly, this formulation, with the participation of the CFJ term, caused the emergence of an unusual mass contribution in the photon Lagrangian density. While the usual Proca mass term breaks only gauge invariance, the present mass contribution breaks Lorentz symmetry and, supposedly, gauge invariance. These result looked intriguing, since the model from which we begin, in which $F_{\mu \nu}$ is treated as an independent field, has not any evidence of violation of gauge symmetry. In this paper, we study a more general version of the massive model obtained in \cite{Palatini} and show that there is a residual gauge invariance in the model obtained from the Palatini-like formulation.
	
	The paper is organized as follows: in section II, general discussions on the model are presented and the gauge field propagator is obtained; in section III, we carry out an analysis of the field equations; we study tree level unitarity of the model in section IV; in section V, we investigate the presence of a residual gauge invariance in the massive model; conclusions and final remarks are presented in section VI.

	\section{General discussions on the model}
	Let us consider the following Lagrangian density of a Lorentz-breaking extension of spinorial quantum electrodynamics (QED),
	\begin{equation}
	{\cal L}=-\frac 14 F_{\mu\nu}F^{\mu\nu} + \alpha \mu \kappa_\mu A_\nu \partial_\beta A_\sigma \varepsilon^{\mu\nu\beta\sigma} 
	+ \frac 12 \mu^2 A^\mu M_{\mu\nu} A^\nu -J_\mu A^\mu + {\cal L}_F,
	\label{model}
	\end{equation}
	with
	\be
	M_{\mu\nu}= \kappa^2 \eta_{\mu\nu} - \rho \kappa_\mu \kappa_\nu,
	\label{mass-tensor}
	\ee
	in which $\alpha$ and $\rho$ are dimensionless constants, $\mu$ is a mass parameter and $\kappa^\mu$ is a background constant pseudovector that causes the violation of Lorentz symmetry. Besides, $F_{\mu\nu}=\partial_\mu A_\nu - \partial_\nu A_\mu$, $J^\mu=e\bar{\psi}\gamma^\mu \psi$ is the conventional matter current and ${\cal L}_F$ is the pure fermionic Lagrangian density. The gauge invariance of the model is apparently explicitly broken due to the presence of the mass term, $(1/2) \mu^2 A^\mu M_{\mu\nu} A^\nu$, which also includes a Lorentz-breaking part. It is worth to note that such a term is usually called nonminimal, since it accommodates nonminimal couplings to the curvature in field theories in curved space-time \cite{Curv-space1}-\cite{Curv-space4}.
	
	This particular LVI extension of QED, in which the background vector of the Carroll-Field-Jackiw (CFJ) term is responsible for the mass tensor, was obtained in \cite{Palatini}, with $\alpha=-1$, $\rho=1$ and $\mu^2=1$. This was accomplished through a Palatini-type procedure, in which the $F^{\mu\nu}$ tensor and the gauge field were initially considered to be independent. The result was an unusual situation, in which the breaking of Lorentz's invariance induces a violation, at least apparent, of the gauge symmetry.
	
	The breaking of gauge symmetry may cause problems to the consistency of the theory, as the violation of unitarity, with deleterious effects in the renormalizability of the model. A common procedure in cases like that is the searching for a hidden gauge symmetry, like the solution of Stueckelberg \cite{Stueckelberg}, in which a mixed term that includes a scalar field is added to the Lagrangian density, such that gauge symmetry is restored. It is noteworthy that the idea of the Stueckelberg mechanism was extended to the Standard Model to give mass to the physical photon through the hypercharge $U(1)_Y$ \cite{Stueckelberg-2}, \cite{Stueckelberg-3}. Another approach is the application of a dualization method in order to obtain a gauge theory which describes the same physics as the original model. In both cases, it is assumed that the model of interest is a gauge-fixed version of a gauge theory. We will perform further investigations on the possibility of finding a hidden gauge symmetry in the present model. Let us first analyse general features of model (\ref{model}).
	
	The quadratic part of the Lagrangian density in the vectorial field $A^\mu$ can be written as
	\be
	{\cal L}_G=\frac 12 A^\mu{\cal O}_{\mu\nu}A^\nu,
	\ee
	with
	\be
	{\cal O}_{\mu\nu}=\left(\Box + \mu^2 \kappa^2\right)\theta_{\mu\nu} + \mu^2 \kappa^2 \omega_{\mu\nu} 
	- 2 \alpha \mu S_{\mu\nu}- \rho \mu^2 \Lambda_{\mu\nu},
	\ee
	where $\theta_{\mu\nu}=\eta_{\mu\nu}-\partial_\mu \partial_\nu/\Box$ and $\omega_{\mu\nu}=\partial_\mu \partial_\nu/\Box$ are the transversal and longitudinal spin projectors, respectively, and the operators
	\be
	S_{\mu\nu}=\varepsilon_{\mu\nu\alpha\beta}\kappa^\alpha \partial^\beta  \,\,\,\,\, \mbox{and} \,\,\,\,\, 
	\Lambda_{\mu\nu}=\kappa_\mu \kappa_\nu
	\ee
	are dependent on the background vector $\kappa^\mu$. The wave operator ${\cal O}_{\mu\nu}$ is invertible. However, as we intend to define a closed algebra which includes the operators $\theta_{\mu\nu}$, $\omega_{\mu\nu}$, $\Lambda_{\mu\nu}$ and $S_{\mu\nu}$, it is necessary to include new operators, as the equation (\ref{14}) indicates. It should be noted that no new term, such as a gauge-fixing, is being added to the wave operator. Since we have
	\be
	S_{\mu \alpha }S_{\,\,\,\,\,\nu }^{\alpha } =(\kappa^{2}\square -\lambda ^{2}) \theta _{\mu \nu } - \lambda ^{2}\omega _{\mu \nu } + \lambda \left( \Sigma _{\mu \nu }+\Sigma _{\nu \mu }\right) - \square \Lambda _{\mu \nu } \equiv f_{\mu\nu },  \label{14}
	\ee
	with
	\begin{equation}
	\Sigma _{\mu \nu }=\kappa_{\mu }\partial _{\nu }\;,\;\;\lambda \equiv \Sigma
	_{\mu }^{\;\mu }=\kappa_{\mu }\partial ^{\mu },
	\end{equation}
	we include the operators $\Sigma_{\mu\nu}$ and $\Sigma_{\nu\mu}$ in the algebra, which is displayed in Table 1. 
	\vspace{2mm}
	
	\begin{widetext}
		\begin{center}
			\begin{tabular}{|c|c|c|c|c|c|c|}
				\hline
				& $\theta _{\,\,\,\,\,\nu }^{\alpha }$ & $\omega _{\,\,\,\,\,\nu }^{\alpha }$
				& $S_{\,\,\,\,\,\nu }^{\alpha }$ & $\Lambda _{\,\,\,\,\,\nu }^{\alpha }$ & $
				\Sigma _{\,\,\,\,\,\nu }^{\alpha }$ & $\Sigma_\nu ^{\,\,\,\,\,\alpha}$ \\ 
				\hline
				$\theta _{\mu \alpha }$ & $\theta _{\mu \nu }$ & $0$ & $S_{\mu \nu }$ & $
				\Lambda _{\mu \nu }-\frac{\lambda }{\square }\Sigma _{\nu \mu }$ & $\Sigma
				_{\mu \nu }-\lambda \omega _{\mu \nu }$ & $0$ \\ \hline
				$\omega _{\mu \alpha }$ & $0$ & $\omega _{\mu \nu }$ & $0$ & $\frac{\lambda 
				}{\square }\Sigma _{\nu \mu }$ & $\lambda \omega _{\mu \nu }$ & $\Sigma_{\nu
					\mu}$ \\ \hline
				$S_{\mu \alpha }$ & $S_{\mu \nu }$ & $0$ & $f_{\mu \nu }$ & $0$ & $0$ & $0$
				\\ \hline
				$\Lambda _{\mu \alpha }$ & $\Lambda _{\mu \nu }-\frac{\lambda }{\square}
				\Sigma_{\mu \nu }$ & $\frac{\lambda }{\square }\Sigma _{\mu \nu }$ & $0$ & $
				\kappa^{2}\Lambda _{\mu \nu }$ & $\kappa^{2}\Sigma _{\mu \nu }$ & $\lambda
				\Lambda_{\mu \nu}$ \\ \hline
				$\Sigma _{\mu \alpha }$ & $0$ & $\Sigma _{\mu \nu }$ & $0$ & $\lambda
				\Lambda _{\mu \nu }$ & $\lambda \Sigma _{\mu \nu }$ & $\square\Lambda_{\mu \nu}$ \\ \hline
				$\Sigma_{\alpha \mu}$ & $\Sigma_{\nu \mu} -\lambda \omega_{\mu \nu}$ & $
				\lambda \omega_{\mu \nu}$ & $0$ & $\kappa^2 \Sigma_{\nu \mu}$ & $\kappa^2 \square
				\omega_{\mu \nu}$ & $\lambda \Sigma_{\nu \mu}$ \\ \hline
			\end{tabular}
			\vspace{2mm}
			
			Table 1: Multiplicative table. The products are supposed to obey the order ``row times
			column''.
		\end{center}
	\end{widetext}
	\vspace{2mm}
	
	We thus obtain, after a straightforward but lengthy calculation, the vector field propagator, $\left<A_\mu A_\nu \right>=i ({\cal O}^{-1})_{\mu\nu}$, in momentum-space:
	\bq
	&&\left<A_\mu A_\nu \right>=\frac{i}{D}\left\{-(p^2 - \mu^2 \kappa^2) \theta_{\mu\nu} + 
	\frac{1}{\mu^2 \kappa^2} \left[D + \frac{\lambda^2}{\tilde{D}}\left(p^2 H + 4\alpha^2 \mu^2 \tilde{D}\right) \right] \omega_{\mu\nu} +
	\right.\nonumber \\
	&& \left.  + 2i \alpha \mu S_{\mu\nu} - \frac{\lambda H}{\tilde D}  (\Sigma_{\mu\nu} + \Sigma_{\nu\mu}) - 
	\frac{\mu^2}{\tilde{D}}\left(-\kappa^2 H + 4 \alpha^2 \tilde{D} \right) \Lambda_{\mu\nu} \right\},
	\eq 
	with
	\be
	D=(p^2-\mu^2 \kappa^2)^2 + 4\alpha^2 \mu^2(\kappa^2 p^2 - \lambda^2),
	\ee
	\be
	\tilde{D}=-(\kappa^2 p^2 - \rho \lambda^2) + \mu^2 \kappa^4(1-\rho)
	\ee
	and
	\be
	H=-\rho(p^2-\mu^2 \kappa^2) + 4\alpha^2 \mu^2 \kappa^2 (1-\rho).
	\ee
	The operators, in momentum-space, are given by
	\bq
	&&\theta_{\mu\nu}=\eta_{\mu\nu} - \omega_{\mu\nu}, \,\,\,\,\omega_{\mu\nu}=\frac{p_\mu p_\nu}{p^2},
	\,\,\,\, S_{\mu\nu}=\varepsilon_{\mu\nu\alpha\beta}\kappa^\alpha p^\beta, \nonumber \\
	&&\Sigma_{\mu\nu}=\kappa_\mu p_\nu \,\,\,\, \mbox{and} \,\,\,\, \Lambda_{\mu\nu}=\kappa_\mu \kappa_\nu.
	\label{op-momentum}
	\eq
	
	We recognize above the propagator for the model of \cite{Palatini}, if we fix $\alpha=-1$, $\rho=1$ and $\mu^2=1$. Interesting that the dispersion relations extracted from the denominator $D$, responsible for the massive modes, are not affected by the constant $\rho$. This parameter only influences the modes coming from $\tilde{D}$. In \cite{Palatini}, the physical consistency of the modes of propagation of the vector field were studied in detail. For now, let us study the effects of the constant $\rho$ on the model and, for comparison, let us fix $\alpha=-1$ and $\mu^2=1$. 
	
	\section{Field equations}
	
	Since in \cite{Palatini} it was shown that only a spacelike $\kappa^\mu$ produces meaningful modes of propagation of the vector field, we restrict ourselves here to the analysis of this case. Let us consider a referential frame in which we have a purely spacelike background vector given by $\kappa^\mu=(0, \vec{\kappa})$, with $|\vec{\kappa}|^2=t^2$, and choose the direction of the $z$-axis as the one towards which the $\vec{\kappa}$ vector is pointed. So, we have the poles 
	\be
	p_0^2=p_\bot^2 +(p_z+t)^2 \equiv m_+^2 \,\,\,\,\,\, \mbox{and} \,\,\,\,\,\, p_0^2=p_\bot^2 +(p_z-t)^2 \equiv m_-^2,
	\ee
	due to the denominator $D$, and
	\be
	p_0^2=p_\bot^2 +(1-\rho)(p_z^2-t^2) \equiv \tilde{m}^2,
	\ee
	due to the denominator $\tilde{D}$, where $\vec{p}_\bot$ is the component of $\bold{p}$ orthogonal to the $z$-axis.
	
	We first examine the solutions of the field equation in momentum-space,
	\be
	(-p^2+\kappa^2)A_\mu + (p \cdot A)p_\mu - \rho (\kappa \cdot A) \kappa_\mu + 
	2i \varepsilon_{\mu\nu\alpha\beta}A^\nu \kappa^\alpha p^\beta = 0,
	\label{field equation}
	\ee
	with $A^\mu=(\phi,\bold{A})$. The contraction of the above equation with $p^\mu$ gives us the gauge condition $(p \cdot A)=\frac{\rho \lambda}{\kappa^2}(\kappa \cdot A)$, which, after the substitution in (\ref{field equation}), furnishes us
	\be
	(-p^2+\kappa^2)A_\mu + \rho \left(\frac{\lambda}{\kappa^2}p_\mu - \kappa_\mu \right) (\kappa \cdot A) + 
	2i \varepsilon_{\mu\nu\alpha\beta}A^\nu \kappa^\alpha p^\beta = 0,
	\ee
	in which $\lambda=\Sigma_\mu^\mu$, according to the definitions of (\ref{op-momentum}) for the operators in momentum-space. For our spacelike $\kappa^\mu$, we have
	\be
	-(p^2+t^2)A_\mu - \rho \left(p_z p_\mu + t^2 \delta_\mu^3 \right) A_z + 
	2it \varepsilon_{\mu\nu 3 \beta}A^\nu p^\beta = 0.
	\ee
	
	In addition, let us define the $x$-axis as the direction aligned with the component of $\bold{A}$ which is orthogonal to $\vec{\kappa}$, so that $A_y=0$. For the $z$-component of the field equation, we have
	\be
	\left[p_0^2-p_\bot^2-(1-\rho)(p_z^2-t^2)\right]A_z=0.
	\ee
	We see that the pole $p_0^2=\tilde{m}^2$ automatically satisfies the equation above. For $A_z \neq 0$, the poles $p_0^2=m_\pm^2$ are constrained to equal $\tilde{m}^2$, which means, for a non null $\rho$,
	\bq
	(p_z + t)^2=(1-\rho)(p_z^2-t^2) \quad \Rightarrow \quad
	p_z &=& \left\{
	\begin{array}{rr}
		-t, &  \mbox{if} \quad 0<\rho \leqslant 1\\
		\left(\frac{\rho-2}{\rho}\right) t, & \mbox{if} \quad \rho > 1 \quad \mbox{or} \quad \rho < 0
	\end{array}
	\right., \quad \mbox{for} \quad \tilde{m}^2=m_+^2 
	\eq
	and
	\bq
	(p_z - t)^2=(1-\rho)(p_z^2-t^2) \quad \Rightarrow \quad
	p_z &=& \left\{
	\begin{array}{rr}
		t, &  \mbox{if} \quad 0 < \rho \leqslant 1	\\
		\left(\frac{2 - \rho}{\rho}\right) t, & \mbox{if} \quad \rho > 1 \quad \mbox{or} \quad \rho < 0 
	\end{array}
	\right., \quad \mbox{for} \quad \tilde{m}^2=m_-^2.
	\label{pz condition}
	\eq
	It is not possible that these two poles satisfy these conditions at the same time for an arbitrary $\rho$, if $t \neq 0$. The other equations of motion for $A_z \neq 0$, with $p_0^2 = \tilde{m}^2$ and, consequently, $p^2+t^2 = -\rho (p_z^2-t^2)$, are written as
	\bq
	\left\{
	\begin{array}{rcr}
		-\rho (p_z^2-t^2)A_x + \rho p_x p_z A_z + 2it p_y \phi &=& 0 \\
		\rho p_y p_z A_z + 2it(p_0 A_x - p_x \phi) &=& 0 \\
		\rho (p_z^2-t^2) \phi - \rho p_0 p_z A_z -2it p_y A_x &=& 0
	\end{array}
	\right.,
	\label{system}
	\eq
	We try a solution with $\phi=0$, for which, after some manipulation of equations (\ref{system}), we obtain
	\be
	(p_0^2-p_y^2)A_x=0.
	\ee
	If $A_x \neq 0$, we are left with the constraints $p_x=0$ and $p_z = \pm t$, which correspond to the two modes $\tilde{m}^2=m_+^2$ ($p_z=-t$) and $\tilde{m}^2=m_-^2$ ($p_z=t$). For the first mode, the electromagnetic field is polarized as $\bold{A}=(A_x,0,A_z)$ and propagates in the direction $\bold{p}=(0,p_y,-t)$, such that $\bold{p} \cdot \bold{A} = -t A_z$ and $p \cdot A = t A_z$. The gauge condition, however, imposes one more restriction:
	\be
	(p \cdot A)=\frac{\rho \lambda}{\kappa^2}(\kappa \cdot A) \quad \Rightarrow \quad (p \cdot A) = \rho t A_z.
	\ee
	So, this solution fixes $\rho = 1$. For the mode with $\tilde{m}^2=m_-^2$ the conclusions are the same, only with the change in the sign of the $p_z$ component.
	
	If $A_x=0$, taking into account the other equations, we necessarily have $p_z=0$, such that $\tilde{m}^2=p_\bot^2-(1-\rho)t^2$. Since, the poles $m_\pm^2$ satisfy the $3$-component of field equations for $p_z=\pm (\rho-2)t/\rho$, we have that, if $\rho=2$, the three poles are reduced to one $p_0^2=\tilde{m}^2=m_\pm^2 = p_\bot^2 + t^2$, with the possible non physical consequences of a multiple pole. 
	
	In the case $A_z=0$, the mode $p_0^2=\tilde{m}^2$ is not excluded. It is not possible, however, to fix $\phi=0$. In order to obtain a solution which is valid for all values of $\rho$, we obtain the constrainings $p_y=0$ and $p_z = \pm t$, which satisfies one of the two conditions: $p_0^2=\tilde{m}^2=m_+^2$ or $p_0^2=\tilde{m}^2=m_-^2$.
	
	In summary, in this section we investigated how the solutions of the field equations are affected by the poles of the model. In the next section, we will look at the physical nature of these poles subject to specific conditions imposed by the field equations.
	
	\section{Tree-level unitarity analysis}
	
	It is enlightening that we make an analysis of the physical nature of the poles. With this purpose, we investigate the tree-level unitarity of the model. It can be investigated through the propagator, when saturated by conserved currents,
	\be
	{\cal SP}=J^\mu \left<A_\mu A_\nu \right> J^{*\nu}.
	\ee
	The current conservation in momentum-space is written as $p_\mu J^\mu=0$, such that
	\be
	p \cdot J = p_0 J_0 - \bold{p} \cdot \bold{J}=0 \quad \Rightarrow \quad J_0 = \frac{\bold{p} \cdot \bold{J}}{p_0}
	\ee
	and
	\be
	J^\mu J^*_\mu = |J_0|^2 - |J_\bot|^2 - |J_z|^2
	= \frac{1}{p_0^2} \left[|\bold{p}\cdot \bold{J}|^2 - p_0^2\left(|J_\bot|^2 + |J_z|^2\right) \right].
	\ee
	Unitarity requires that the imaginary part of the residue of the saturated propagator in a physical pole is nonnegative (see \cite{Cas-Man} and \cite{Veltman}). This requirement can be checked by calculating the residue matrix in the pole for the complete propagator and, then, verifying if its eigenvalues are nonnegative. Here we opt to analyze directly the saturated propagator. Note that in \cite{Cas-Man} this technique of analysis was used in a model obtained from the dimensional reduction of the CPT-even sector of the Standard Model Extension. 
	
	Turning our attention to the saturated propagator, in consequence of current conservation, only terms on $\eta_{\mu\nu}$ and $\Lambda_{\mu\nu}$ remains, such that, for our spacelike $\kappa^\mu$, we stay with
	\be
	{\cal SP}=\frac{i}{p_0^2 D}\left\{-(p^2 + t^2) \left[|\bold{p}\cdot \bold{J}|^2 - p_0^2|J_\bot|^2 \right] + \frac{p_0^2 |J_z|^2}{\tilde{D}}\left[-t^4 H + (p^2 - 3t^2) \tilde{D}\right]\right\},
	\ee
	with
	\be
	D=(p_0^2-m_+^2)(p_0^2-m_-^2),
	\ee
	\be
	\tilde{D}=t^2(p_0^2-\tilde{m}^2)
	\ee
	and
	\be
	H=-\rho(p^2+t^2)-4t^2(1-\rho).
	\ee
	
	We have some potential problematic situations to analyze here: when $p_z=\pm t$, we have, in principle, double poles, with $\tilde{m}^2=m_+^2$ and $\tilde{m}^2=m_-^2$, respectively; this situation repeats when $p_z = \pm \left(\frac{\rho-2}{\rho}\right)t$, with an additional complication 
	if $\rho=2$, with $p_z=0$, which, at first sight, brings us a triple pole ($\tilde{m}^2=m_+^2=m_-^2$). We study below these cases.
	
	\subsection{The cases $p_z = \pm t$}
	Since these two cases are similar, we analyze here the referential frame in which $p_z=-t$, such that we stay with the poles $p_0^2=\tilde{m}^2=m_+^2= p_\bot^2$ and $p_0^2 = m_-^2 = p_\bot^2 + 4t^2$. This situation is possible both for $A_z=0$ and $A_z \neq 0$, as we verified in the field equations. The key point in order to analyze the possibility of a double pole is the fact that $p^2+t^2 = p_0^2 - \tilde{m}^2$ in this referential frame. We stay with
	\bq
	&&{\cal SP}=\frac{i}{p_0^2(p_0^2-\tilde{m}^2-4t^2)}\left\{-\left[ |\bold{p} \cdot \bold{J}|^2 - p_0^2 |J_\bot|^2\right] \right. \nonumber \\
	&& \left. + \frac{p_0^2 |J_z|^2}{(p_0^2-\tilde{m}^2)^2} \left[(p_0^2-\tilde{m}^2)\left(p_0^2 -\tilde{m}^2 + (\rho - 4)t^2  \right) + 4t^4(1-\rho)\right]\right\}
	\eq
	and we see that, in order to eliminate the double pole, we need to have $\rho=1$, in accordance with the result we obtained in the study of the field equations in conjunction with the gauge condition. If we fix $\rho = 1$, the saturated propagator reads
	\be
	{\cal SP}=\frac{i}{p_0^2(p_0^2-\tilde{m}^2-4t^2)}\left\{-\left[|\bold{p}\cdot \bold{J}|^2 - p_0^2 |J_\bot|^2 \right] + \frac{p_0^2 |J_z|^2}{(p_0^2-\tilde{m}^2)}\left(p_0^2 -\tilde{m}^2 - 3t^2  \right) \right\}.
	\ee
	
	We are now in position to calculate the imaginary part of the residue in the two remaining poles. For the pole $p_0^2=\tilde{m}^2$, the calculation is straightforward and give
	\be
	\mbox{Im}\left[{\cal R}_{p_0^2=\tilde m^2}({\cal SP})\right]=\frac 34 |J_z|^2.
	\ee
	For the residue in the pole $p_0^2=m_-^2$, the term in $|\bold{p}\cdot \bold{J}|^2$ contributes. The residue gives us
	\be
	\mbox{Im}\left[{\cal R}_{p_0^2=p_\bot^2 + 4 t^2}({\cal SP})\right] = \frac{1}{4(\tilde{m}^2 + 4 t^2)}\left\{-4 |\bold{p}\cdot \bold{J}|^2 + 4(\tilde{m}^2 + 4 t^2) |J_\bot|^2 + (\tilde{m}^2 + 4 t^2) |J_z|^2 \right\}
	\ee
	We then remember that, for $p_z=-t$, we have $p_x=0$ and that, in this case, $\tilde{m}^2=p_y^2$ to obtain, after some calculation,
	\be
	\mbox{Im}\left[{\cal R}_{p_0^2 = p_y^2 + 4 t^2}({\cal SP})\right] =  \frac{|p_y J_z + 4tJ_y|^2}{4(p_y^2 + 4t^2)} + |J_x|^2 > 0.
	\ee
	The result above, which can be diagonalized, indicates a massive mode with two degrees of freedom.
	
	\subsection{The cases $p_z = \pm \left(\frac{\rho-2}{\rho}\right)t$}
	We here analyze the case $p_z = \left(\frac{\rho-2}{\rho}\right)t$, since the two situations are similar. In this case, we have a possible double pole at 
	\begin{equation}
	p_0^2=\tilde{m}^2=m_+^2 = p_\bot^2 +  \frac{4(1-\rho)^2t^2}{\rho^2},  
	\end{equation}
	and, if $\rho=2$, a triple pole. The saturated propagator, after algebraic manipulations, is given by
	\bq
	&& {\cal SP}= \frac{i}{p_0^2(p_0^2-\tilde{m}^2)\left[p_0^2-\tilde{m}^2 - \frac{4(2-\rho)t^2}{\rho}\right]} \left\{ -\left[p_0^2-\tilde{m}^2 + \frac{4(\rho-1)t^2}{\rho}\right]
	\left(|\bold{p}\cdot \bold{J}|^2 - p_0^2 |J_\bot|^2 \right) \right. \nonumber \\
	&& \left. + p_0^2 |J_z|^2 \left[p_0^2-\tilde{m}^2 + \frac{(\rho^2-4)t^2}{\rho}\right] \right\}
	\eq
	The residue in the pole $p_0^2=\tilde{m}^2$ gives
	\begin{equation}
	\mbox{Im}\left[ {\cal R}_{p_0^2 = \tilde{m}^2} ({\cal SP}) \right]= \frac{(\rho + 2)}{4}|J_z|^2 + \frac{(1-\rho)}{(2-\rho)}\frac{1}{\tilde{m}^2} \left[|\bold{p}\cdot \bold{J}|^2 - \tilde{m}^2 |J_\bot|^2 \right].
	\end{equation}
	
	It is clear that a nonnegative imaginary part of the residue in the pole $p_0^2=\tilde{m}^2$ depends on the value of $\rho$. In the case of $\rho=1$, we are back to the case $p_z=-t$. The possible triple pole for $\rho=2$ and $p_z=0$ is not consistent, since the residue diverges. This physical inconsistency of such a case is also related to the stability of the model and will be made clear in the next section.

	\section{The residual gauge-invariance}
	All the above results present us clues of some special feature in the model when $\rho=1$. First, let us rewrite the dispersion relations for the model in a more general way. From the $D$ factor in the denominator, taking $\alpha=1$ and $\mu=1$, we have
	\begin{equation}
	(p^2-\kappa^2)^2 + 4(\kappa^2 p^2 - \lambda^2) = 0,
	\end{equation}
	which can be factorized as
	\begin{equation}
	(p^2 + \kappa^2 + 2 \lambda)(p^2 + \kappa^2 - 2 \lambda)=0,
	\end{equation}
	so that
	\begin{equation}
	\omega^2 \equiv p_0^2 = |\bold{p}|^2 - \kappa^2 \pm 2 \lambda.
	\end{equation}
	For the purely spacelike background vector, $\kappa^\mu=(0,\vec{\kappa})$, we stay with
	\begin{equation}
	\omega^2= \left| \bold{p} \pm \vec{\kappa} \right|^2.
	\end{equation}
	Since the model is massive, we can go to the particle rest frame and get, for these two dispersion relations,
	\begin{equation}
	\omega^2 = |\vec{\kappa}|^2,
	\end{equation}
	which represents a mode with a positive definite mass which propagates with two degrees of freedom. It is important to emphasize that these results do not depend on the value of the $\rho$ parameter. On the other hand, the $\tilde{D}$ factor is written as
	\begin{equation}
	\tilde{D}= \rho \lambda^2 - \kappa^2 p^2 + (1-\rho) \kappa^4,
	\end{equation}
	which furnishes us the dispersion relation
	\begin{equation}
	\omega^2 = |\bold{p}|^2 + \frac{1}{\kappa^2}\left[\rho \lambda^2 + (1-\rho) \kappa^4  \right].
	\end{equation}
	For the purely spacelike Lorentz-breaking vector, the result is written as
	\begin{equation}
	\omega^2 = |\bold{p}|^2 - \frac{1}{|\vec{\kappa}|^2}\left[ \rho (\bold{p} \cdot \vec{\kappa})^2 + (1-\rho) | \vec{\kappa}|^4 \right] = |\bold{p}|^2 \left(1 - \rho \cos^2 \theta  \right) - (1-\rho) |\vec{\kappa}|^2,
	\end{equation}
	where $\theta$ is the angle between the vectors $\bold{p}$ and $\vec{\kappa}$. Let us consider the field propagates in the direction of the Lorentz-breaking vector, such that $\theta=0$. In this situation, we stay with
	\begin{equation}
	\omega^2=(1-\rho)\left[|\bold{p}|^2 - |\vec{\kappa}|^2 \right],
	\end{equation}
	which imposes some restrictions. First, if $\rho=1$, we have a massless mode. For $\rho \neq 1$, the dispersion relation predicts a massive mode, but with the restriction $\rho > 1$, otherwise we have an unstable model, since in the particle rest frame we have $\omega^2=(\rho - 1)|\vec{\kappa}|^2$. However, even with $\rho > 1$ the stability of the model imposes severe restrictions, as the magnitude of the field momentum would be limited by that of the background vector, which is very small. The propagation of the field would be practically restricted to the plane orthogonal to $\vec{\kappa}$, although the component of the momentum in the direction of $\vec{\kappa}$ cannot be null (we must have $\rho \neq 2$). Thus, in the special case in which $\rho = 1$ the model presents a massive mode, $\omega^2 = |\vec{\kappa}|^2$ (coming from the factor $D$ in the denominator), with two degrees of freedom, and a massless mode, which corresponds to a propagation in the direction of the background vector (from $\tilde{D}$). 
	
	In massive models with violation of gauge symmetry, it is usual to look for hidden symmetries through dualization processes, which could provide an equivalent gauge-invariant model. The gauge embedding method \cite{Ilha}, \cite{NDM} proved to be effective in this type of procedure in a wide range of models. The approach is based on the transformation of the model in a gauge theory by adding on mass-shell vanishing terms. This iterative embedding of Noether counterterms is based on the idea of local lifting a global symmetry and is reminiscent to procedures which were important in the construction of component-field supergravity actions \cite{sugra}. It would be interesting to carry out the dualization of the present theory. However, the procedure needs the use of the inverse operator of the mass tensor, $M_{\mu \nu}$ (defined in equation (\ref{mass-tensor})), which is given by
	\begin{equation}
	L_{\mu \nu} = \frac{1}{\kappa^2}\left\{\eta_{\mu \nu} + \rho \frac{\kappa_\mu \kappa_\nu}{(1-\rho)\kappa^2} \right\}.
	\end{equation}
	Interestingly, there is no inverse for $M_{\mu\nu}$ when $\rho=1$. This is another compelling clue for the investigation of this particular case.
	
	Last but not least, calculating the one-loop vacuum polarization tensor for the model up to second-order in $\kappa^\mu$ gives us a transverse result \cite{Palatini}, with
	\begin{equation}
	\Pi_{\mu\nu}^{(2)}= -\frac{ie^2}{6\pi^2m^2}\left[1+\frac 25 \frac{p^2}{m^2} 
	+ {\cal O}\left(\frac{p^4}{m^4}\right) \right] T_{\mu\nu},
	\end{equation}
	in which $m$ is the fermion mass and
	\begin{equation}
	T_{\mu\nu}= \kappa^2(p_\mu p_\nu - p^2 \eta_{\mu\nu}) + p^2\kappa_\mu \kappa_\nu +(p\cdot \kappa)^2 \eta_{\mu\nu} - (p \cdot \kappa)(p_\mu \kappa_\nu + p_\nu \kappa_\mu). 
	\end{equation}
	
	All these results bring us the possibility of a residual gauge-invariance. The gauge-violating term is given by
	\begin{equation}
	{\cal L}_M= \frac 12 A^\mu M_{\mu\nu} A^\nu.
	\end{equation}
	By considering a gauge transformation on the vectorial field, $A^\mu$,
	\begin{equation}
	A^\mu \to A^\mu + \partial^\mu \chi,
	\end{equation}
	we have
	\begin{equation}
	\delta {\cal L} = A^\mu M_{\mu \nu} (\partial^\nu \chi) + \frac 12 (\partial^\mu \chi) M_{\mu \nu} (\partial^\nu \chi).
	\end{equation}
	It is easy to see that if $\partial^\mu \chi$ is proportional to $\kappa^\mu$, such that $\partial^\mu \chi = \beta(z) \kappa^\mu$, with $\beta(z)$ a dimensionless function of $z$ (assuming the $z$-axis is parallel to the background vector), we get, explicitly using (\ref{mass-tensor}),
	\begin{equation}
	(\partial^\mu \chi) M_{\mu\nu}= \beta \kappa^2 \kappa_\nu (1 - \rho),
	\end{equation}
	which is null for $\rho=1$. In this situation, the variation of the Lagrangian is null. Just in order to illustrate, for our purely spacelike vector $\kappa^\mu=(0,\vec{\kappa})$, this is accomplished if the gradient of $\chi$ is parallel to $\vec{\kappa}$ in the case $\rho=1$. Thus, there is a class of gauge transformations which leaves the Lagrangian density invariant and, in this sense, we can affirm that the model accommodates a residual gauge-invariance when $\rho=1$.
	
	Although we use an specific model in the discussion of the residual gauge-invariance, it is easy to see that this is a general result for all models whose only source of gauge symmetry violation is the above cited ${\cal L}_M$.

	\section{Conclusion}
	\label{sec-conclusion}
	We studied an alternative CPT-odd electrodynamics which incorporates a general Lorentz-breaking mass term. Actually, it is an extended version of the one obtained from a Palatini-like procedure in \cite{Palatini}. We have showed that the Lorentz-breaking part of the mass tensor affects only one of the three modes of propagation of the gauge-field. The $\rho$ parameter that controls this part of the mass tensor plays a fundamental role in preserving the essential physical properties of the model. It was shown, for a spacelike Lorentz-breaking vector $\kappa^\mu$,  that if $\rho < 1$, the model presents serious problems with stability. Moreover, if this parameter is greater than one, a positive definite energy requires the field is restricted to propagate in a direction external to a cone defined by the rotation, around the $z$-axis (defined as the direction of the background vector), of the lines $z = \pm (t/p_\bot) u$, being $u$ the axis parallel to the component $\vec{p}_\bot$ of the field momentum and $t$ the magnitude of the Lorentz-breaking vector $\vec{\kappa}$. Moreover, the particular case in which $\rho=2$ accommodates a nonphysical triple pole when the component of the field momentum along the $z$-axis is null.
	
	The most interesting case, however, is the one with $\rho=1$, which presents a massless mode of propagation along the $z$-axis. This fact, together with the lack of an inverse for the mass tensor, led us to investigate the presence of a residual gauge-invariance in the model. We then showed that a gauge transformation, $A^\mu \to A^\mu + \partial^\mu \chi$, such that $\chi=\chi(z)$, leaves the action unchanged. For a purely spacelike background vector, this corresponds to have the gradiant of $\chi$ parallel to $\vec{\kappa}$. Moreover, this is a general result for models whose only source of gauge symmetry violation is the mass term studied in the present paper.
	
	For a future work, it would be interesting to investigate how a dualization process would work in the limit $\rho \to 1$. In other words, it would be worth to understand what kind of model comes up in this limit in a more general gauge-invariant model with $\rho$ undetermined, obtained from a gauge embedding procedure.

	\vspace{1.0cm}
	
	\noindent {\bf Acknowledgments}
	
	This work was partially supported by Conselho Nacional de Desenvolvimento Cient\'{\i}fico e Tecnol\'{o}gico (CNPq). The authors acknowledge Prof. J. A. Helay\"el-Neto for elucidating discussions.
	


\begin{thebibliography}{99}
		
		\bibitem{qgrav1} V.A.~Kosteleck\'y, S.~Samuel, Phys. Rev. D \textbf{39}, 683 (1989);
		V.A.~Kosteleck\'y and S.~Samuel, Phys. Rev. D \textbf{40}, 1886 (1989).
		
		
		\bibitem{qgrav2} S.~Doplicher, K.~Fredenhagen, J.E.~Roberts,
		Commun. Math. Phys.  \textbf{172}, 187 (1995);
		J.~Collins, A.~Perez, D.~Sudarsky, L.~Urrutia, H.~Vucetich,
		Phys. Rev. Lett. \textbf{93}, 191301 (2004);
		P.~Horava,
		Phys. Rev. D \textbf{79}, 084008 (2009).
		
		\bibitem{CFJ} S.M.~Carroll, G.B.~Field,  R.~Jackiw,
		Phys. Rev. D \textbf{41}, 1231 (1990).
		
		
		\bibitem{CFJ1} A.P. Ba\^eta Scarpelli, H. Belich, J.L. Boldo, J.A. Helay\"el-Neto, Phys. ¨
		Rev. D 67, 085021 (2003); A.P. Ba\^eta Scarpelli and J.A. Helay\"el-Neto, ¨
		Phys.Rev. D 73, 105020 (2006).
		
		\bibitem{CFJ2} R. Jackiw and V.A. Kosteleck\`y, Phys. Rev. Lett. 82, 3572 (1999); J.-M.
		Chung and B.K. Chung, Phys. Rev. D 63, 105015 (2001); J.-M. Chung,
		Phys. Rev. D 60, 127901 (1999); G. Bonneau, Nucl. Phys. B 593, 398
		(2001); M. Perez-Victoria, Phys. Rev. Lett. 83, 2518 (1999); J. High.
		Energy Phys. 0104, 032 (2001); O.A. Battistel and G. Dallabona, Nucl.
		Phys. B 610, 316 (2001); J. Phys. G 27, L53 (2001); J. Phys. G 28, L23
		(2002); A.P. Ba\^eta Scarpelli, M. Sampaio, M.C. Nemes, and B. Hiller, ˆ
		Phys. Rev. D 64, 046013 (2001);
		A.P. Ba\^eta Scarpelli, M. Sampaio, M.C. Nemes, B. Hiller, Eur. Phys. J. C \textbf{56}, 571 (2008).
		
		\bibitem{CFJ3} C. Adam and F.R. Klinkhamer, Nucl. Phys. B 607, 247 (2001); Nucl.
		Phys. B 657, 214 (2003).
		
		
		\bibitem{CFJ4} A.A. Andrianov and R. Soldati, Phys. Rev. D 51, 5961 (1995); Phys. Lett.
		B 435, 449 (1998); A.A. Andrianov, R. Soldati, and L. Sorbo, Phys. Rev.
		D 59, 025002 (1998); A.A. Andrianov, D. Espriu, P. Giacconi, and R. Soldati, J. High Energy Phys. 0909, 057 (2009); J. Alfaro, A.A. Andrianov,
		M. Cambiaso, P. Giacconi, and R. Soldati, Int. J. Mod. Phys. A 25, 3271
		(2010); V.Ch. Zhukovsky, A.E. Lobanov, and E.M. Murchikova, Phys.
		Rev. D 73, 065016, (2006).
		
		\bibitem{SME1} D. Colladay and V. A. Kostelecky, Phys. Rev. D 55, (1997) 6760.
		
		\bibitem{SME2} D. Colladay and V. A. Kostelecky, Phys. Rev. D 58, (1998) 116002.
		
		\bibitem{SME3} S. Coleman and S. L. Glashow, Phys. Lett. B 405, (1997) 249;
		S. Coleman and S. L. Glashow, Phys. Rev. D 59, (1999) 116008.
		
		\bibitem{data} V.A. Kosteleck\`y, N. Russell, Rev. Mod. Phys. 83, 11 (2011).
		
		
		\bibitem{Proca1}   P. Robles, F. Claro,
		Eur. J. Phys. \textbf{33}, 1217 (2012).
		
		\bibitem{Proca2}   L.C. Tu, J. Luo, G.T.~Gillies,
		Rept. Prog. Phys. \textbf{68}, 77 (2005).
		
		\bibitem{Proca3} A.S. Goldhaber, M.M. Nieto, Rev. Mod. Phys. \textbf{82}, 939 (2010).
		
		\bibitem{Proca4} G. Spavieri, J. Quintero, G.T. Gillies, M. Rodriguez, Eur. Phys. J. D \textbf{61}, 531 (2011).
		
		\bibitem{Proca5} M.V.S. Fonseca, A.V. Paredes, Braz. J. Phys \textbf{40}, 319 (2010).
		
		\bibitem{mass-limit-1}   J.-J. Wei, X.-F. Wu,
		JCAP \textbf{1807}, 045 (2018);
		L.~Shao, B.~Zhang,
		Phys. Rev. D \textbf{95}, 123010 (2017);
		L.~Bonetti, J.~Ellis, N.E.~Mavromatos, A.S.~Sakharov, E.K.~Sarkisyan-Grinbaum, A.D.A.M.~Spallicci,
		Phys. Lett. B \textbf{768}, 326 (2017);
		Y.-P.~Yang, B.~Zhang,
		Astrophys. J.  \textbf{842}, 23 (2017);
		J.-J.~Wei, E.-K.~Zhang, S.-B.~Zhang, X.-F.~Wu,
		Res. Astron. Astrophys.  \textbf{17}, 13 (2017);
		B.~Zhang, Y.-T.~Chai, Y.-C.~Zou, X.-F.~Wu,
		JHEAp \textbf{11-12}, 20 (2016);
		L.~Bonetti, J.~Ellis, N.E.~Mavromatos, A.S.~Sakharov, E.K.G.~Sarkisyan-Grinbaum, A.D.A.M.~Spallicci,
		Phys. Lett. B \textbf{757}, 548 (2016);
		X.-F.~Wu {\it et al.},
		Astrophys. J. \textbf{822}, L 15 (2016).
		
		\bibitem{mass-limit-2} A. Retin\`o, A.D.A.M. Spallicci, A. Vaivads, Astropart. Phys. \textbf{82}, 49 (2016).
		
		\bibitem{mass-limit-3} M. Tanabashi et al. (Particle Data Group), Phys. Rev. D \textbf{98}, 030001 (2018).
		
		\bibitem{mass-limit-4} D.D. Ryutov, Plasma Phys. Contr. Fus. \textbf{39}, A 73 (1997).
		
		\bibitem{mass-limit-5} D.D. Ryutov, Plasma Phys. Contr. Fus. \textbf{49}, B 429 (2007).
		
		\bibitem{mass-limit-6} L. Davis, A.S. Goldhaber,  M.M. Nieto, Phys. Rev. Lett. \textbf{35}, 1402 (1975).
		
		\bibitem{mass-limit-7} G. Chibisov, Usp. Fiz. Nauk \textbf{19}, 551 (1976).
		
		\bibitem{Bonetti-1} L.~Bonetti, L.R.~dos Santos Filho, J.A.~Helay\"el-Neto, A.D.A.M.~Spallicci,
		Phys. Lett. B \textbf{764}, 203 (2017).
		
		\bibitem{Bonetti-2} L. Bonetti, L.R. dos Santos, J.A. Helay\"el-Neto, A.D.A.M. Spallicci, Eur. Phys. J. C \textbf{78}, 811 (2018).
		
		\bibitem{Manoel-Helayel} Manoel M. Ferreira, Jose A. Helay\"el-Neto, Carlos M. Reyes, Marco Schreck, Pedro D.S. Silva, Phys. Lett. B \textbf{804}, 135379 (2020).
		
		\bibitem{LVmass1} G. Gabadadze, L. Grisa, Phys. Lett. B \textbf{617}, 124 (2005).
		
		\bibitem{LVmass2} G. Dvali, M. Papucci, M.D. Schwartz, Phys. Rev. Lett. \textbf{94}, 191602 (2005).
		
		\bibitem{LVmass3} B. Altschul, Phys. Rev. D \textbf{73}, 036005 (2006).
		
		\bibitem{LVmass4} M. Cambiaso, R. Lehnert, R. Potting, Phys. Rev. D \textbf{85}, 085023 (2012).
		
		\bibitem{LVmass5} B. Altschul, Phys. Rev. D \textbf{86}, 045008 (2012).
		
		\bibitem{LVmass6} H.G. Fargnoli, L.C.T. Brito, A.P. Ba\^eta Scarpelli, M. Sampaio, Phys. Rev. D \textbf{90}, 085016 (2014).
		
		\bibitem{Palatini} J.C.C. Felipe, H.G. Fargnoli, A.P. Ba\^eta Scarpelli, L.C.T. Brito, Int.J.Mod.Phys.A 34, n 25, 1950139, e-Print: 1807.00904 [hep-th].
		
		\bibitem{Curv-space1} D.J. Toms, \textit{Quantization of the minimal and non-minimal vector field in
			curved space}, arXiv:1509.05989
		
		\bibitem{Curv-space2} I.L. Buchbinder, T.P. Netto, I.L. Shapiro, Phys. Rev. D \textbf{95}, 085009 (2017).
		
		\bibitem{Curv-space3} M.S. Ruf, C.F. Steinwachs, Phys. Rev. D \textbf{98}, 025009 (2018).
		
		\bibitem{Curv-space4} C. Garcia-Recio, L.L. Salcedo, Eur. Phys. J. C \textbf{79}, 438 (2019).
		
		\bibitem{Stueckelberg} E.C.G. Stueckelberg, Helv. Phys. Acta \textbf{11}, 225 (1938);
		E. C. G. Stueckelberg, Helv. Phys. Acta \textbf{11}, 299 (1938).
		
		\bibitem{Stueckelberg-2} Boris K\"ors, Pran Nath, Phys. Lett. B \textbf{586}, 366 (2004).
		
		\bibitem{Stueckelberg-3} Henri Ruegg, Mart\'i Ruiz-Altaba, Int. J. Mod. Phys. A \textbf{19}, 3265 (2004).
		
		\bibitem{Cas-Man} R. Casana, M.M. Ferreira Jr., R.P.M. Moreira, Eur. Phys. J. C \textbf{72}, 2070 (2012).
		
		\bibitem{Veltman} M. Veltman, ``Quantum Theory of Gravitation'', in ``Methods in Field Theory'', edited by R. Bailian and J. Zinn-Justin
		(North-Holland Publising Company and World Scientific Publising Co Ltd, Singapore, 1981).
		
		\bibitem{Ilha} A. Ilha, C. Wotzasek, Nucl. Phys. B \textbf{604}, 426 (2001)
		
		\bibitem{NDM} M. A. Anacleto, A. Ilha, J. R. S. Nascimento, R. F. Ribeiro, C. Wotzasek, Phys. Lett. B \textbf{504}, 268 (2001).
		
		\bibitem{sugra} D. Z. Freedman and P. van Nieuwenhuizen, Phys. Rev. D \textbf{13}, 3214 (1976);
		S. Ferrara, D. Z. Freedman and P. van Nieuwenhuizen, Phys. Rev. D \textbf{15}, 1013 (1977);
		S. Ferrara and J. Scherk, Phys. Rev. Lett. \textbf{37}, 1035 (1976).
		
		
		
	\end{thebibliography}
\end{document}